\documentclass[twoside,twocolumn,9pt]{article}
\usepackage{extsizes}
\usepackage[super,sort&compress,comma]{natbib} 
\usepackage[version=3]{mhchem}
\usepackage[left=1.5cm, right=1.5cm, top=1.785cm, bottom=2.0cm]{geometry}
\usepackage{balance}
\usepackage{mathptmx}
\usepackage{mathrsfs}
\usepackage{sectsty}
\usepackage{graphicx} 
\usepackage{lastpage}
\usepackage{tabularx}
\usepackage{booktabs}
\usepackage{multirow}
\usepackage[format=plain,justification=justified,singlelinecheck=false,font={stretch=1.125,small,sf},labelfont=bf,labelsep=space]{caption}
\usepackage{float}
\usepackage{fancyhdr}
\usepackage{fnpos}
\usepackage[english]{babel}
\addto{\captionsenglish}{%
  
}
\usepackage{array}
\usepackage{droidsans}
\usepackage{charter}
\usepackage[T1]{fontenc}
\usepackage[usenames,dvipsnames]{xcolor}
\usepackage{setspace}
\usepackage[compact]{titlesec}
\usepackage{hyperref}
\usepackage{amssymb}
\usepackage{svg}
\usepackage{threeparttable}

\usepackage{epstopdf}

\definecolor{cream}{RGB}{222,217,201}

\begin{document}

\pagestyle{fancy}
\thispagestyle{plain}
\fancypagestyle{plain}{
\renewcommand{\headrulewidth}{0pt}
}

\makeFNbottom
\makeatletter
\renewcommand\LARGE{\@setfontsize\LARGE{15pt}{17}}
\renewcommand\Large{\@setfontsize\Large{12pt}{14}}
\renewcommand\large{\@setfontsize\large{10pt}{12}}
\renewcommand\footnotesize{\@setfontsize\footnotesize{7pt}{10}}
\makeatother

\renewcommand{\thefootnote}{\fnsymbol{footnote}}
\renewcommand\footnoterule{\vspace*{1pt}%
\color{cream}\hrule width 3.5in height 0.4pt \color{black}\vspace*{5pt}} 
\setcounter{secnumdepth}{5}

\makeatletter 
\renewcommand\@biblabel[1]{#1}            
\renewcommand\@makefntext[1]%
{\noindent\makebox[0pt][r]{\@thefnmark\,}#1}
\makeatother 
\renewcommand{\figurename}{\small{Fig.}~}
\sectionfont{\sffamily\Large}
\subsectionfont{\normalsize}
\subsubsectionfont{\bf}
\setstretch{1.125} 
\setlength{\skip\footins}{0.8cm}
\setlength{\footnotesep}{0.25cm}
\setlength{\jot}{10pt}
\titlespacing*{\section}{0pt}{4pt}{4pt}
\titlespacing*{\subsection}{0pt}{15pt}{1pt}

\fancyfoot{}
\fancyfoot[LO,RE]{\vspace{-7.1pt}\includegraphics[height=9pt]{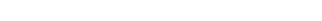}}
\fancyfoot[CO]{\vspace{-7.1pt}\hspace{13.2cm}\includegraphics{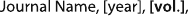}}
\fancyfoot[CE]{\vspace{-7.2pt}\hspace{-14.2cm}\includegraphics{head_foot/RF}}
\fancyfoot[RO]{\footnotesize{\sffamily{1--\pageref{LastPage} ~\textbar  \hspace{2pt}\thepage}}}
\fancyfoot[LE]{\footnotesize{\sffamily{\thepage~\textbar\hspace{3.45cm} 1--\pageref{LastPage}}}}
\fancyhead{}
\renewcommand{\headrulewidth}{0pt} 
\renewcommand{\footrulewidth}{0pt}
\setlength{\arrayrulewidth}{1pt}
\setlength{\columnsep}{6.5mm}
\setlength\bibsep{1pt}

\makeatletter 
\newlength{\figrulesep} 
\setlength{\figrulesep}{0.5\textfloatsep} 

\newcommand{\topfigrule}{\vspace*{-1pt}%
\noindent{\color{cream}\rule[-\figrulesep]{\columnwidth}{1.5pt}} }

\newcommand{\botfigrule}{\vspace*{-2pt}%
\noindent{\color{cream}\rule[\figrulesep]{\columnwidth}{1.5pt}} }

\newcommand{\dblfigrule}{\vspace*{-1pt}%
\noindent{\color{cream}\rule[-\figrulesep]{\textwidth}{1.5pt}} }

\makeatother

\twocolumn[
  \begin{@twocolumnfalse}
\vspace{1em}
\sffamily
\begin{tabular}{m{4.5cm} p{13.5cm} }

\includegraphics{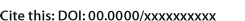} & \noindent\LARGE{\textbf{DiffBindFR: An SE(3) Equivariant Network for Flexible Protein-Ligand Docking$^\dag$}} \\
\vspace{0.3cm} & \vspace{0.3cm} \\

 & \noindent\large{Jintao Zhu,$^{\dag}$\textit{$^{a}$} Zhonghui Gu,$^{\dag}$\textit{$^{b}$} 
 Jianfeng Pei,\textit{$^{\ddag a}$} and Luhua Lai\textit{$^{\ddag a,b,c}$}}\\

\includegraphics{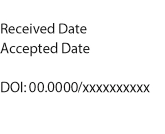} & \noindent\normalsize{Molecular docking, a key technique in structure-based drug design, plays pivotal roles in protein-ligand interaction modeling, hit identification and optimization, in which accurate prediction of protein-ligand binding mode is essential. Conventional docking approaches perform well in redocking tasks with known protein binding pocket conformation in the complex state. However, in real-world docking scenario without knowing the protein binding conformation for a new ligand, accurately modeling the binding complex structure remains challenging as flexible docking is computationally expensive and inaccurate. Typical deep learning-based docking methods do not explicitly consider protein side chain conformations and fail to ensure the physical plausibility and detailed atomic interactions. In this study, we present DiffBindFR, a full-atom diffusion-based flexible docking model that operates over the product space of ligand overall movements and flexibility and pocket side chain torsion changes. We show that DiffBindFR has higher accuracy in producing native-like binding structures with physically plausible and detailed interactions than available docking methods. Furthermore, in the Apo and AlphaFold2 modeled structures, DiffBindFR demonstrates superior advantages in accurate ligand binding pose and protein binding conformation prediction, making it suitable for Apo and AlphaFold2 structure-based drug design. DiffBindFR provides a powerful flexible docking tool for modeling accurate protein-ligand binding structures.} \\



\end{tabular}

 \end{@twocolumnfalse} 
 \vspace{0.6cm}

  ]

\renewcommand*\rmdefault{bch}\normalfont\upshape
\rmfamily
\section*{}
\vspace{-1cm}


\footnotetext{\textit{$^{a}$~Center for Quantitative Biology, Academy for Advanced Interdisciplinary Studies, Peking University, Beijing, 100871, China.}}
\footnotetext{\textit{$^{b}$~Peking-Tsinghua Center for Life Sciences, Academy for Advanced Interdisciplinary Studies, Peking University, Beijing, 100871, China. }}
\footnotetext{\textit{$^{c}$~BNLMS, College of Chemistry and Molecular Engineering, Peking University, Beijing, 100871, China.}}

\footnotetext{\dag~ These authors contributed equally to this work.}
\footnotetext{\ddag~ To whom correspondence should be addressed.}




\section{Introduction}
The primary paradigm of drug discovery involves identifying and designing molecules that target key proteins within disease pathways. Historically, screening compound libraries using biochemical platforms has been the predominant approach for identifying novel drugs~\citep{handen2002high}. Since the 1990s, high-throughput screening (HTS) has been employed on libraries ranging from 500,000 to 10$^{8}$ molecules~\citep{mayr2009novel,satz2022dna}, leading to the discovery of several drugs. While the HTS libraries represent a significant advancement over traditional lab-designed ones, they encompass only a fraction of potential drug-like molecules~\citep{fink2005virtual}. Given the challenges and expenses associated with synthesizing such a vast chemical space, computational methods for screening virtual libraries are frequently employed in drug discovery, allowing exploration of chemical spaces comprising tens of billions of molecules or even more~\citep{grebner2019virtual,sadybekov2023computational,lyu2023modeling}.

Structure-based virtual screening (SBVS) enables rapid and cost-effective modeling of target-molecule binding structures from large-scale compound libraries together with the evaluation of their binding affinities for identifying potential hits~\citep{gorgulla2020open,zhu2022comprehensive,lyu2019ultra}. Molecular docking is one of the most frequently employed techniques for SBVS, which is utilized to predict ligand binding poses, characterize protein-ligand binding strength, and identify key interactions~\citep{fan2019progress,bender2021practical}. In general, conventional docking approaches, including AutoDock4~\citep{morris2009autodock4}, AutoDock Vina~\citep{trott2010autodock,eberhardt2021autodock}, Smina~\citep{koes2013lessons}, Glide~\citep{friesner2004glide}, and GOLD~\citep{jones1997development}, leverage heuristic search algorithms, to explore a variety of potential ligand conformations. Scoring functions with simplified terms are utilized for fast estimation of binding affinity and priority of ligand poses.

\begin{figure*}[!h]
\centering
\includegraphics[scale=0.07]{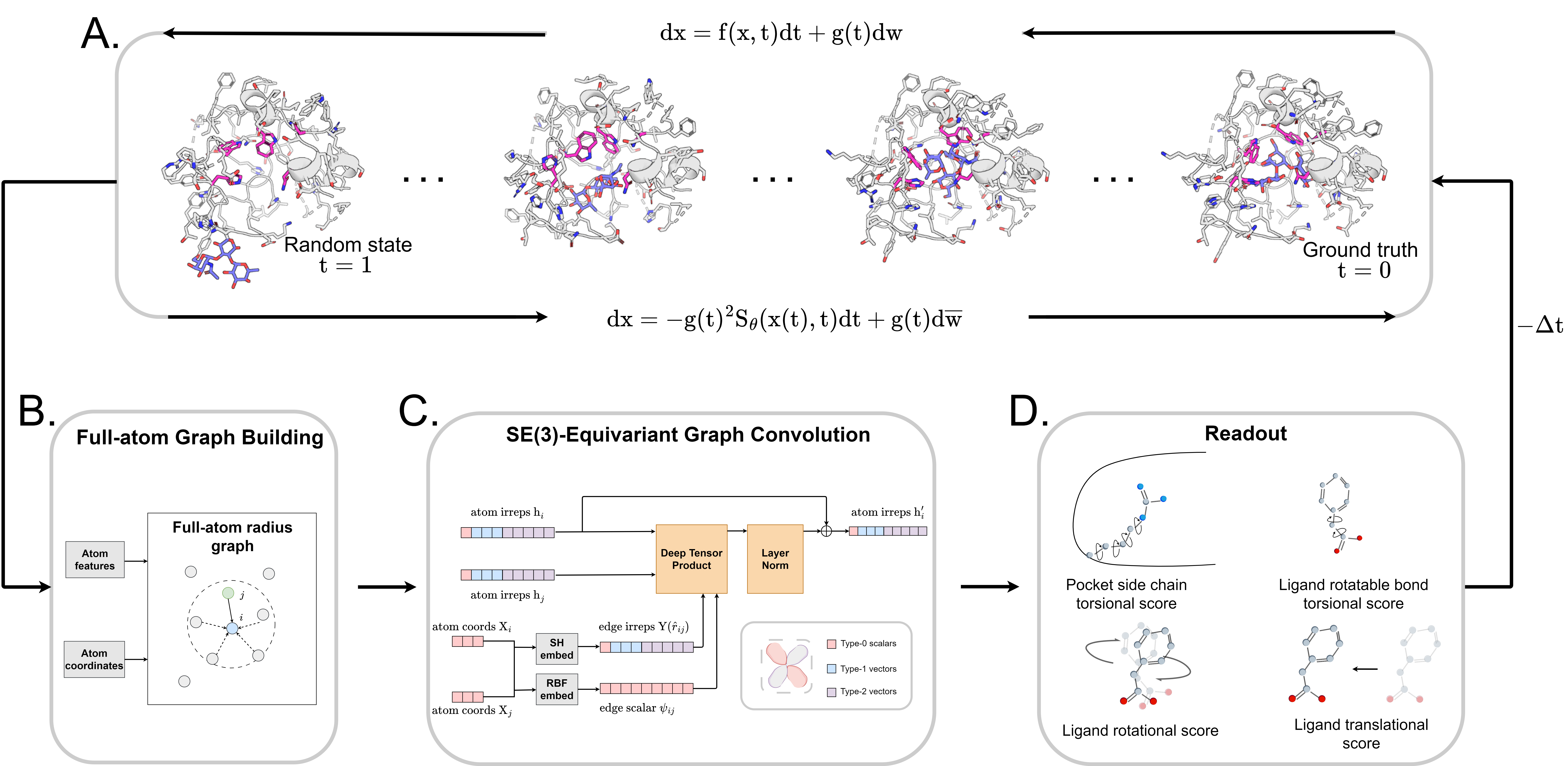}
\caption{The architecture of DiffBindFR. (A). Overview of score-based generative modeling through SDE for flexible docking. The flexible docking process is decomposed into ligand translation, rotation, bond torsion and pocket side chain torsion. (B). Construction of full-atom interaction graph. According to the real-time coordinate of each atom, we build the spatial graph as model input. (C). The architecture of SE(3) equivariant graph convolution. It serves as the trunk block of DiffBindFR network. $\mathrm{h_i}$ and $\mathrm{X_i}$ are the irreducible representations (Irreps) and coordinate of atom $\mathrm{i}$, respectively. The distance and vector between atom $\mathrm{i}$ and atom $\mathrm{j}$ are embedded through Gaussian radial basis (RBF) and spherical harmonics basis (SH) respectively, to get their edge scalar representations $\mathrm{\psi_{ij}}$ and edge vector Irreps $\mathrm{Y(\hat{r}_{ij})}$. Then, Deep Tensor Product from e3nn library is served as message-passing module to gather messages from neighborhood, followed by an equivariant Layer Normalization (Layer Norm) module to get the updated Irreps $\mathrm{h_i^\prime}$. (D). The output readout of DiffBindFR network contains the predicted score of pocket side chain torsion, ligand rotatable  bond torsion, ligand rotation and ligand translation. These scores are used to solve the reverse SDE for binding structures sampling.}

\label{fig:1}
\end{figure*}

Classical molecular docking methods describe protein-ligand interactions based on the lock-and-key model~\citep{lauria2011virtual}, wherein a rigid receptor binding pocket serves as the "lock" and the molecular docking algorithm primarily optimizes the ligand's conformation to find a complementary "key". Such rigid receptor docking methods, for the trade-off between accuracy and computational efficiency, strive to determine the optimal and complementary binding conformation. When known complex structures are available, and ligand molecules are removed and then re-docked into the native Holo pockets, rigid receptor docking often achieves impressive success rate~\citep{wang2016comprehensive}. However, in real-world docking tasks without knowing the binding conformation in advance, ligand induced pocket conformational changes may produce wrong docking results~\citep{antunes2015understanding}. Virtual screening and rational design on the unbound (Apo) or computationally modeled structures usually give unsatisfied hit rate~\citep{mcgovern2003information,lee2009improving,zhang2022holo,diaz2023deep,zhang2023benchmarking,kersten2023hic}. Without accounting for pocket flexibility, the performance of docking methods experiences drastic decrease in such cases~\citep{scior2012recognizing}, which may rule out potential hits during the early stage of drug discovery. Although AlphaFold2~\citep{jumper2021highly} is capable of accurately modeling target protein structures, traditional docking methods that overlook potential side chain flexibility perform less effectively when applied to the predicted structures\citep{karelina2023accurately}. 

Currently, there are two main strategies to address the flexibility of protein pockets. The first approach involves inducing local conformational rearrangements of the target using force field or scoring function-based calculations. For instance, rDock~\citep{ruiz2014rdock} allows movements of functional groups that can form hydrogen bonds including -OH and -NH3. AutoDockFR (AutoDock for Flexible Receptors)~\citep{ravindranath2015autodockfr} allows users to specify up to 14 flexible side chains in advance and samples reasonable side chain dihedral angles from a rotamer library. Despite its better performance than AutoDock Vina in cross-docking benchmarks with Apo structures, it is considerably time-consuming and requires prior knowledge of potentially critical side chains in the pocket, which limits its application in SBVS. The second approach is the recently developed deep learning-based methods~\citep{yu2023deep}, which coarsen the representation of protein pockets by only encoding the protein backbone atoms without explicitly including side chain atoms. This kind of representation is insensitive to minor pocket backbone flexibility and side chain adaptability. Earlier works, like DeepDock~\citep{mendez2021geometric}, TankBind~\citep{lu2022tankbind}, and EDM-Dock~\citep{masters2023deep}, predicted pocket residue-ligand distance map, which is used to reconstruct the binding structure. Leveraging powerful equivariant neural networks like EGNN~\citep{satorras2021n}, geometric deep learning~\citep{atz2021geometric} models such as EquiBind~\citep{stark2022equibind}, LigPose~\citep{Zhang2022ligpose}, E3Bind~\citep{zhang2022e3bind}, Uni-Mol~\citep{zhou2022uni}, and KarmaDock~\citep{zhang2023efficient} iteratively predict the three-dimensional coordinates of ligands directly around the whole protein (blind docking) or predefined pocket. Recent SOTA blind docking method DiffDock~\citep{corso2022diffdock}, based on the diffusion generative modelling~\citep{ho2020denoising}, employed the SE(3) equivariant neural network~\citep{geiger2022e3nn} to denoise the rotation, translation, and bond torsion of ligand, and then rank poses by additional confidence model. 
However, these existing deep learning-based docking approaches face limitations in effectively handling protein flexibility and the generated ligand poses are often implausible~\citep{buttenschoen2023posebusters}. The generated ligand structures often contain clashes with the target and irrational bond lengths, angles, and torsion angles that lead to high intra energies. Ligand conformational optimization using tools such as RDKit alignment~\citep{zhang2023efficient,riniker2015better} cannot completely alleviate ligand and protein clashes. Furthermore, ignoring the target flexibility and validity of ligand poses makes it challenging for these deep learning-based methods to capture key interactions in docking~\citep{buttenschoen2023posebusters}. Recently, building upon the methodology of optimizing ligand coordinate recycling as developed in LigPose and KarmaDock, a deep learning-based flexible docking method named FlexPose~\citep{dong2023equivariant} has extened its predictive capabilities of pocket side chain coordinates. This advancement allows for more details of interaction information in cross-docking applications. However, like LigPose and KarmaDock, FlexPose encounters the same inherent limitation. The methodological focus on fitting coordinates within Euclidean space tends to overfit the overall RMSD (Root Mean Square Deviation). Consequently, FlexPose, akin to its predecessors, is inevitably limited by conformational rationality. 
Overall, these inadequacies of current methods impede subsequent steps, such as post-optimization of ligands by experts based on the detailed interactions or conducting further studies through molecular dynamics simulations. 

Early Apo-Holo pair analysis has shown general consensus that upon ligand binding protein pocket undergoes significant side chain conformation heterogeneity while backbone is relatively rigid in most cases~\citep{gaudreault2012side,clark2019inherent,wankowicz2022ligand}. Therefore, in most cases, side chain flexibility modelling is enough for flexible docking. In this study, we developed a full-atom flexible docking model, DiffBindFR, based on the diffusion framework (Fig.\ref{fig:1}). In the comprehensive evaluation, starting from pocket conformations with randomized side chain torsion angles, DiffBindFR outperforms state-of-the-art (SOTA) deep learning techniques and traditional docking methods. Owing to the explicit full-atom modeling of pocket residues, and its learning of joint optimization of variables within the entire system across a product space composed of torsional angles, rotations, and translations, DiffBindFR can not only accurately recover protein pocket side chain conformations, but also generate precise and highly physically plausible ligand binding poses. In cross-docking benchmark, DiffBindFR, due to its capability of full pocket side chain optimization, significantly outperforms traditional flexible docking methods like AutoDock VinaFlex~\citep{ravindranath2015autodockfr} and rDock~\citep{ruiz2014rdock}. It also gives superior performance in generating both accurate and valid binding structures compared to existing docking methods in the Apo dataset and AlphaFold2 modeled structures.

\section{Overview of DiffBindFR}
We formulated flexible docking as a problem of learning the joint denoising process of four variables in their tangent space: ligand rotation $\mathrm{R}$, translation $\mathrm{T}$, rotatable bond torsion $\mathrm{\tau}$, and pocket side chain torsion $\mathrm{\chi}$. Following the VE-SDE (variance exploding stochastic differential equation) paradigm~\citep{song2020score}, starting from the crystal complex $\mathrm{P(x(0)) = P(R(0),T(0),\tau(0),\chi(0))}$, the forward process of the diffusion model, $\mathrm{P(x(t)|x(0))}$, involves uniformly and continuously sampling time step $\mathrm{t \in [0,1]}$ and injecting noise to the four kinds of movement operator to achieve binding structure perturbation. DiffBindFR is an SE(3)-equivariant generative model, following the message-passing paradigm~\citep{gilmer2017neural} of graph neural network, that encodes the intricate interactions between the full-atom pocket and ligand, and predicts the scores $\mathrm{\nabla_{x(t)}logP_t(x(t))}$~\citep{song2020score}. In the docking procedure, starting from the randomly initialized binding conformation, the scores predicted by DiffBindFR are used to solve the reverse VE-SDE process~\citep{song2020score} to implement denoising sampling. With physics-based scoring function Smina~\citep{koes2013lessons} or mixture density neural network (MDN)~\citep{mendez2021geometric} serving as confidence model, binding structures sampled by DiffBindFR can be ranked, and then the top-1 complex pose can be selected as the final prediction.

\subsection{Diffusion generative model}
The diffusion model utilize the framework of stochastic differential equations~\citep{anderson1982reverse} to diffuse the data distribution described as follows:
\begin{equation}
\mathrm{dx = f(x,t)dt + g(t)dw}
\end{equation}
For $\mathrm{x \in \mathbb{R}^D}$, $\mathrm{{f(x,t) \in \mathbb{R}^{D \times D}}}$ denotes a vector-valued function called the drift coefficient of $\mathrm{x(t)}$, and $\mathrm{g(t) \in \mathbb{R}^{R \times R}}$ denotes a scalar function called the diffusion coefficient of $\mathrm{x(t)}$. The lack of canonical local coordinate system defined for ligand molecules, makes the drift coefficient hard to design for the ligand rotation.
Consequently, the drift coefficient $\mathrm{f(x,t)}$ is set to be $\mathrm{0}$, and the model becomes the score-based generative model~\citep{song2020score}. The reverse diffusion running backwards in time, which is also known as the denoising process, is given by the following reverse-time SDE:
\begin{equation}
\mathrm{dx = - g(t)g(t)^T \nabla_{x(t)}logP_t(x)dt + g(t)d\overline{w}}
\end{equation}
To estimate $\mathrm{\nabla_{x(t)}logP_t(x)}$, we can train a score-based neural network $\mathrm{S_\theta(x(t),t)}$ to fit it. The standard score-match loss function is as follows:
\begin{equation}
\begin{split}
\mathrm{J(x) = \mathbb{E}_t[\lambda(t)\mathbb{E}_{x(t) \sim P_{t|0}(x(t)|x(0))}[||S_\theta(x(t),t)} \\
\mathrm{- \nabla_{x(t)}logP_{t|0}(x(t)|x(0))||^2]]}
\end{split}
\end{equation}
$\mathrm{\lambda(t) = 1 / \mathbb{E}_{x(t)\sim P_{t|0}(x(t)|x(0))}[||\nabla_{x(t)}logP_{t|0}(x(t)|x(0))||^2]}$ is the precomputed weight factors.

\subsection{Pose transformations and diffusion on the product space}
We choose the specific SDE for forward diffusion process as follows:
\begin{equation}
\mathrm{dx = \sqrt{\frac{d\sigma^2(t)}{dt}}dw,\ where\ \sigma(t) = \sigma_{min}^{1-t} \sigma_{max}^{t}, t \in [0,1]}
\end{equation}
$\mathrm{\sigma(t) = \{\sigma_R(t),\sigma_T(t),\sigma_\tau(t),\sigma_\chi(t)\}}$ denotes the noises that injected into ligand Rotation $\mathrm{R}$, translation $\mathrm{T}$, rotatable  bond torsion $\mathrm{\tau}$ and pocket side chain torsion $\mathrm{\chi}$. According to the specific group that each variable lies in, we would design the form of corresponding $\sigma$ carefully for diffusion kernel and the score computation. For a ligand pose with $\mathrm{n}$ atoms, $\mathrm{X_l \in \mathbb{R}^{3 \times n}}$, translation of a ligand pose $\mathrm{T \in \mathbb{R}^3}$ lies in the 3D translation group $\mathbb{T}(3)$. The diffusion kernel for ligand is a Gaussian function with variance $\sigma_T$ as follows, which is also utilized for computing the score of ligand translation $\mathrm{\nabla P_{t|0}(X_l(t)|X_l(0))}$:
\begin{equation}
\mathrm{P_{t|0}(X_l(t)|X_l(0)) =} \mathscr{N}\mathrm{(X_l(0),\sigma_T(t))}
\end{equation}
As rotation of a ligand pose lies in the 3D rotation group $\mathbb{SO}(3)$, $IGSO(3)$ distribution~\citep{nikolayev1997normal,leach2022denoising} was chosen as the diffusion kernel. In specific, rotation matrix $\mathrm{R \in \mathbb{R}^{3 \times 3}}$ can be split into a unit vector $\mathrm{\hat{\omega} \in \mathfrak{so}(3)}$ uniformly as the rotation axis and a axis-angle $\mathrm{\omega \in [0,\pi]}$. Consequently, the functionality of $\mathrm{\sigma_R}$ can be replaced by $\mathrm{\hat{\omega}}$ and $\mathrm{\omega}$. The diffusion kernel for ligand rotation is as follows:
\begin{equation}
\mathrm{P_{t|0}(R(t)|R(0)) = \mathbf{R}(\hat{\omega},\omega)R(0)}
\end{equation}
The score of rotation diffusion can be computed according to
\begin{equation}
\mathrm{\nabla lnP_t(R(t)|R(0)) = (\frac{d}{d\omega}logf(\omega))\hat{\omega}}
\end{equation}

\begin{equation}
\mathrm{f(\omega) = \sum_{l=0}^{\infty}(2l+1)exp(-l(l+1)\epsilon^2/2)\frac{sin((l+1/2)\omega)}{sin(\omega/2)}}
\end{equation}
where $\mathrm{\epsilon}$ is a scalar variance for parameterizing the $IGSO(3)$ distribution.
Torsion of pocket side chains and ligand rotatable  bonds lie in the $SO(2)^m$ group and $SO(2)^k$ group respectively, where $\mathrm{m}$ and $\mathrm{k}$ denote the number of all $\chi$ from the pocket side chain and all $\tau$ from the ligand. Since each torsion angle coordinate lies in $[0, 2\pi)$, the m torsion angles of a conformer define a hypertorus $\mathrm{\mathbb{T}^m}$. We introduced the diffusion kernel from the work of Torsional Diffusion~\citep{jing2022torsional} to satisfy angle periodicity, and compute its score $\mathrm{\nabla P_{t|0}(\chi(t)|\chi(0))}$ as follows:
\begin{equation}
\mathrm{P_{t|0}(\chi(t)|\chi(0)) \propto \sum_{d\in\mathbb{Z}^m} exp(-\frac{||\chi(0)-\chi(t)+2\pi d||^2}{2\sigma_\chi^2(t)})}
\end{equation}
Torsion of ligand rotatable  bonds are dealt with the same way as pocket side chains.

Following the equation(3), the loss function is set as follows:
\begin{equation}
J(x) = J(R) + J(T) + \sum_1^k J(\tau) + \sum_1^m J(\chi)
\end{equation}

The forward diffusion and reverse diffusion are both performed in the product space~\citep{rodola2019functional} of $\mathbb{T}(3) \times SO(3) \times SO(2)^k \times SO(2)^m$, corresponding to the aforementioned four kinds of transformation.

During the forward diffusion process, we would sample $t \in [0,1]$ for each pocket-ligand pair, and then utilize the defined diffusion kernel to sample each transformation. The torsions of ligand and pocket side chains are first applied to the pose, followed by translation and rotation.

The starting point of the denoising stage is a ligand conformation generated by RDKit~\citep{riniker2015better} and pocket side chains, with each type of transformation sampling from their $\mathrm{\sigma_{max}}$. According to equation(2), we update complex pose using the predicted score for each type of transformation. After applying the translation and rotation matrix constructed from predicted score, torsion angles get updated. It is noteworthy that there exists entanglement between ligand translation/rotation and its bond torsion, ligand pose need to be re-aligned to the pose before bond torsion, which will lead to model-unaware structural alignment error. With the sampled pocket side chains fixed, we perform fast local energy relaxation on the ligand by Smina ~\citep{koes2013lessons} for error correction, obtaining the final binding conformations. The number of the denoising steps is defined as 22, and 40 poses are sampled for each pocket-ligand pair, which takes in average 50 s when the batch size is set to 16 on a single 32 GB NVIDIA Tesla V100-SXM2 GPU card.

\subsection{Confidence model}
We have explored two approaches to rank the poses generated by DiffBindFR. First, the traditional scoring function Smina is utilized to quickly score the generated full-atom pocket-ligand poses. Second, a deep learning-based scoring model based on mixture density network (MDN) is trained to fit the distance distribution between ligand atoms and pocket residues. The architecture of our MDN model is similar to the scoring module of KarmaDock, and it shares the similar input representations with DiffBindFR. To better cater for the full-atom complex system, we set the distance pairs as each ligand atom with their nearest atoms from each pocket residues. More details of MDN model can be found in Supplementary Section 3.


\begin{figure*}[!h]
\centering
\includegraphics[scale=0.15]{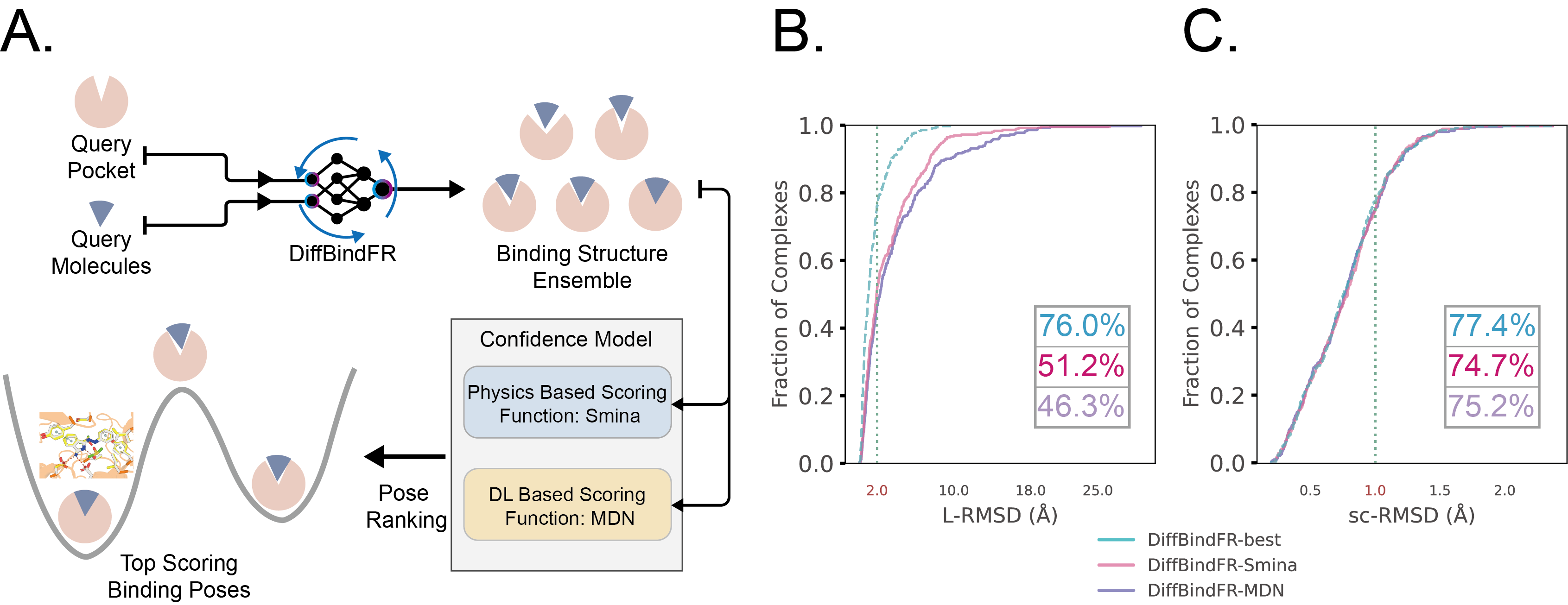}
\caption{(A). Overview of the DiffBindFR workflow. Various complex poses are generated by DiffBindFR network and confidence models are utilized to select the top-1 complex pose. Performance of DiffBindFR in PDBbind time-split test set for L-RMSD (B) and sc-RMSD (C). For each complex, 40 poses are generated. Distributions of L-RMSD and sc-RMSD are computed between DiffBindFR generated poses and ground-truth complex poses. Here, "DiffBindFR-best" means certain metrics are from the pose with the lowest L-RMSD generated by DiffBindFR model; "DiffBindFR-Smina" represents the DiffBindFR generated top-1 poses for each complex ranked by Smina scoring function; "DiffBindFR-MDN" represents the DiffBindFR generated top-1 poses for each complex ranked by MDN confidence model.}

\label{fig:2}
\end{figure*}

\section{Datasets and evaluation metrics}
\subsection{Datasets}
\subsubsection{PDBbind time-split dataset}
We use the PDBbind v2020 dataset~\citep{liu2017forging} for training and evaluation. For each target protein-ligand pair within the PDBbind v2020 dataset, we define the protein pocket as any residues within 12 \AA\ buffer of any heavy atoms in the ligand molecule. Following the time split strategy proposed by the work of EquiBind~\citep{stark2022equibind}, where 363 complex structures uploaded later than 2019 serve as test set. After removing ligands that exist in the test set, the remaining 16739 structures are used for training and 968 structures are used for validation. The dataset is named as "PDBbind time-split dataset" in the article.

The time-split of PDBbind is supported to be more strict and reasonable with the protein sequence similarity of 0.484 between test set and training\&validation set, compared to CASF2016-split whose protein sequence similarity is 1.00 (Supplemental Table. S2). Volkov et al.~\citep{volkov2022frustration} have shown that the time-split of PDBbind is more practical and critical over artificial splits such as ligand scaffolds or protein sequence/structure similarity for the model generalization in drug repurposing, lead optimization, and virtual screening.

\subsubsection{Posebusters test set}
The PoseBusters test set~\citep{buttenschoen2023posebusters} is a meticulously curated collection of crystal complexes sourced from the PDB~\citep{berman2000protein}. This set encompasses a diverse array of high-caliber, recent protein-ligand complexes characterized by drug-like molecules. With 428 distinct complexes, inclusive of unique proteins and ligands released since 2021, it ensures no overlap with the complexes found in the PDBbind v2020 dataset.

\subsubsection{CD test set}
Given the current absence of a large-scale benchmark dataset for cross-docking, especially to address various cross-docking scenarios (including Apo-Holo and cross-docking between different Holo states), we have established a benchmark dataset tailored for the cross-docking evaluation, termed CD test set. We integrated ApoRef~\citep{zhang2022holo}, a test set constructed by constrained MD for inducing Apo-like pockets into Holo-like pockets; several prominent ensemble docking targets including CDK2, EGFR, FXA; CASF2016~\citep{liu2017forging} and GPCR-AF2~\citep{karelina2023accurately} that contains 18 human GPCR complexes published after April 30, 2018. ApoBind database~\citep{aggarwal2021apobind} and AHoJ~\citep{feidakis2022ahoj} are utilized to search for corresponding Apo states based on queried protein-ligand pairs, thereby creating pairs for the Apo-Holo and Holo-Holo mixed cross-dock dataset. The detailed protocol for constructing the CD test set can be found in Supplementary Section 1.
The finalized CD test set comprises of 14,194 structural pairs for cross-docking benchmark tests.

\subsection{Evaluation metrics}
We utilize the Ligand Root Mean Square Deviation (L-RMSD) to assess the predictive quality of ligand conformations. Meanwhile, the evaluation of side chain conformations' predictive quality is based on the side chain Root Mean Square Deviation (sc-RMSD). Let $\mathrm{X_l}^{pred}$ represents the generated ligand pose, and $\mathrm{X_l}^{gt}$ denotes the native ligand pose. 

\subsubsection{L-RMSD}
We take into account Ligand Root Mean Square Deviation (L-RMSD) corrected for symmetry. The precise calculation formula is given below. Herein, $\mathrm{N}$ represents the number of heavy atoms in the ligand, and $\mathrm{isom}$ denotes the isomers of the ligand molecular graph.
\begin{equation}
    \mathrm{L\text{-}RMSD = argmin_{X_l^{isom} \sim isom(X_l^{gt})}\sqrt{\frac{1}{N}\sum_{i=1}^N(X_l^{isom}(i) - X_l^{pred}(i))^2}}
\end{equation}

\subsubsection{Success rate}
L-RMSD $\textless$ 2 \AA\ is widely recognized as a benchmark indication of successful docking~\citep{alhossary2015fast}. In fact, for cross-docking evaluations, the threshold for determining docking success can be relaxed to 2.5 \AA. Nonetheless, to ensure equitable comparison, we adhere to the stricter threshold in this context.

\subsubsection{PB-success rate}
The PoseBusters test suite serves as a rigorous validation tool, assessing both the chemical and geometric consistency of a ligand, inclusive of its stereochemistry. Moreover, it evaluates the physical plausibility of intra-molecular and intermolecular measurements, focusing on factors like the planarity of aromatic rings, canonical bond lengths, and potential protein-ligand clashes. Therefore, the PoseBusters suite provides users a more accurate and realistic estimation of the success rate, PB-success rate, through further checking the physical plausibility of poses with L-RMSD $\textless$ 2 \AA.


\subsubsection{sc-RMSD}
Given that the pocket backbone remains fixed, we compute the RMSD for the side chains of each residue individually and subsequently average the results. Furthermore, in consideration of the symmetrical topology inherent in side chain structures, symmetry corrections have been implemented for the ASP, GLU, PHE, and TYP residues. We regard an sc-RMSD value of less than 1 \AA\ as indicative of success.


\section{Results and discussion}

\begin{figure*}[!h]
\centering
\includegraphics[scale=0.18]{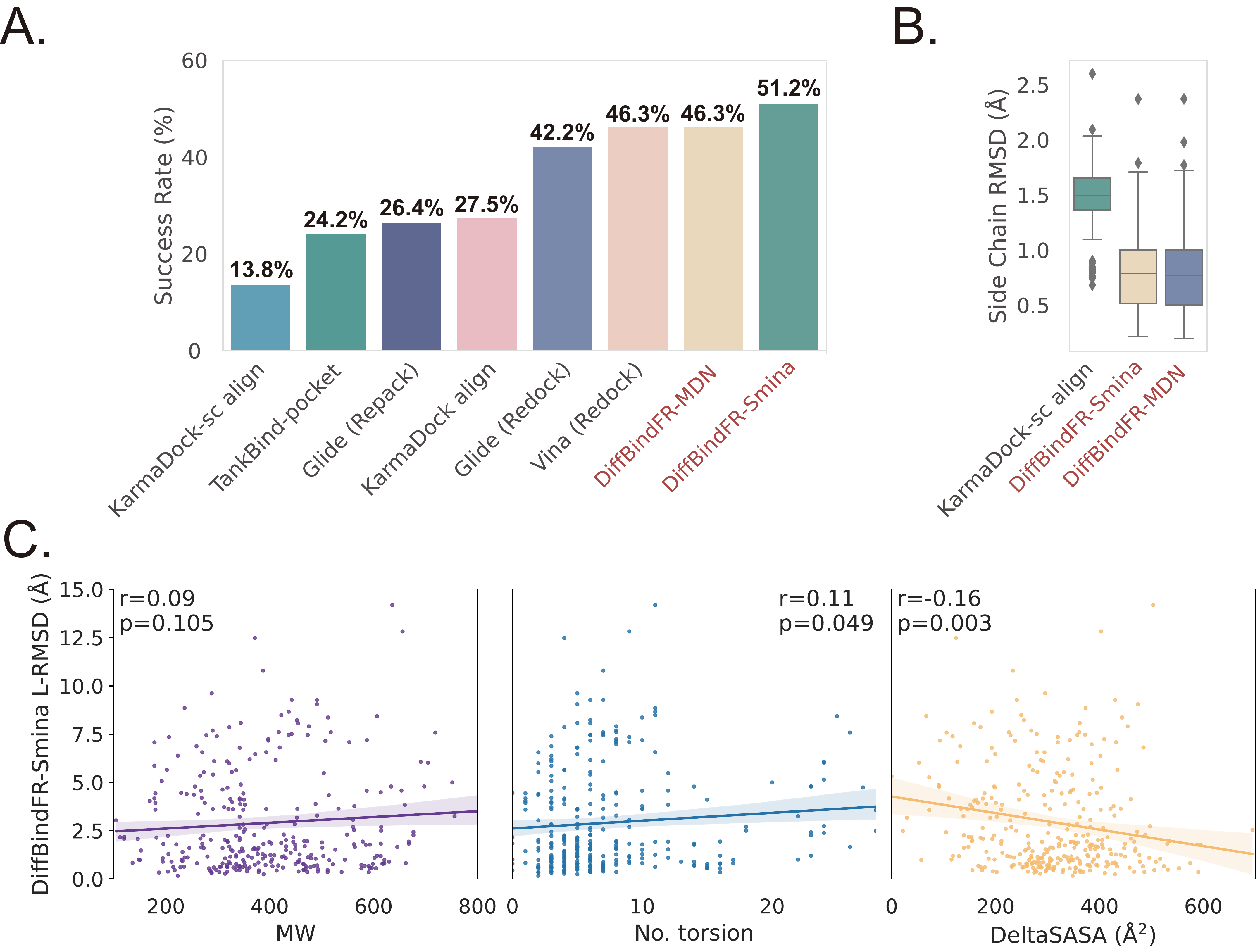}
\caption{(A). Success rate of various pocket docking methods. (B). The distribution of sc-RMSD of KarmaDock-sc Align, DiffBindFR-MDN and DiffBindFR-Smina. (C). The correlation between L-RMSD from DiffBindFR-Smina and molecular weight (MW), number of rotatable  ligand bonds (No. torsion) and variation of solvent accessible surface area caused by binding (DeltaSASA).}

\label{fig:3}
\end{figure*}

\subsection{Performance on the PDBbind time-split test set}
The performance of DiffBindFR is first assessed on the PDBbind time-split test set~\citep{stark2022equibind, liu2017forging}. We employed two metrics, including ligand Root Mean Square Deviation (L-RMSD) and side chain Root Mean Square Deviation (sc-RMSD) for the flexible docking evaluation. As is depicted from Fig.\ref{fig:2}.(B) (DiffBindFR-best), among the best poses (with lowest L-RMSD) from 40 DiffBindFR-generated poses for each complex, 76.0$\%$ of the ligand poses achieve successful docking (L-RMSD $\textless$ 2 \AA). Of these successful DiffBindFR-best poses, 77.4$\%$ exhibit reliable side chain recovery (sc-RMSD \textless 1 \AA). The results of ablation experiments conducted on the hyperparameters related to network sampling and denoising within the DiffBindFR framework are detailed in Supplemental Fig. S3. DiffBindFR-best represents the optimal scenario achievable by the DiffBindFR model, assuming a perfect confidence model could identify all the best poses. To enhance pose selection, we have developed two confidence models for ranking the generated poses (Fig.\ref{fig:2}.(A)). The first model (DiffBindFR-Smina) employs the all-atom physics-based scoring function Smina~\citep{koes2013lessons}, and the second one (DiffBindFR-MDN) utilizes a MDN (Mixture Density Network) network (Fig. S2) trained on the PDBbind time-split training set by this work. Using Smina to select the top-1 binding poses, DiffBindFR-Smina attains a success rate of 51.2$\%$ in the PDBbind time-split test set. Among the top-1 poses ranked by Smina, 74.7$\%$ have reliable side chain recovery. The results demonstrate that DiffBindFR can accurately reconstruct the side chain conformations consistent with experimental pocket-ligand interactions, allowing the Smina to effectively select high-quality binding poses. When ranking sampled poses via the MDN model, DiffBindFR-MDN achieves a little bit lower success rate of 46.3$\%$ compared to DiffBindFR-Smina. Among the top-1 poses ranked by MDN confidence model, 75.2$\%$ have reliable side chain recovery.

\begin{figure*}[!h]
\centering
\includegraphics[scale=0.23]{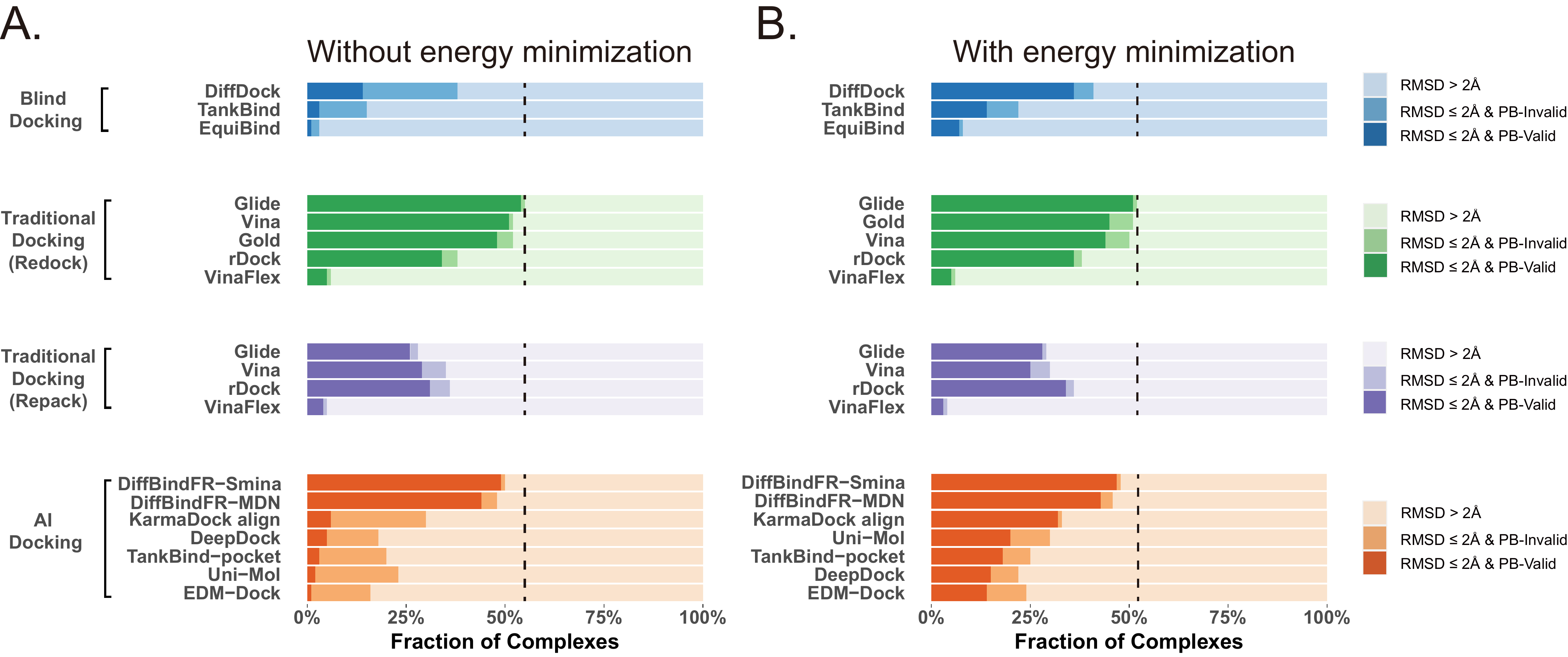}
\caption{Comparison of DiffBindFR with other method boosted by force field optimization in Posebusters test set. The left column (A) shows the performance of methods without energy minimization, and the right one (B) shows the performance of methods with energy minimization. The lightest color represents the failure rate for docking, the moderate color represents the success rate, and the darkest color represent the PB-success rate. The blue color, green color, purple color and orange color represents performance of blind docking methods, traditional docking methods in redocking, traditional docking methods using repacked target proteins as input and deep learning-based docking methods, respectively.}

\label{fig:4}
\end{figure*}

Subsequently, the performance of traditional and recent deep learning-based methods is evaluated for comparison. The top-1 docking poses of each method, selected based on its confidence model or scoring function, are analyzed. DiffBindFR-Smina and DiffBindFR-MDN significantly outperform other deep learning-based pocket docking methods (Fig.~\ref{fig:3}.(A)), including KarmaDock with RDKit~\citep{riniker2015better} ligand conformation alignment (KarmaDock Align) and TankBind with predefined pocket (TankBind-Pocket). This highlights the advantage of our full-atom based model. Even when compared to the traditional rigid receptor docking methods AutoDock Vina and Glide, with the experimentally determined side chain conformations (redock), DiffBindFR-Smina achieves a marginally higher success rate without knowing the side chain conformations in the complex. We further used Rosetta~\citep{rohl2004protein} to repack side chain conformations in these Holo structures to simulate an Apo-like state for each target protein. The docking success rate of Vina and Glide significantly decreases in these Apo-like proteins, underscoring the limitations of rigid receptor docking methods in handling side chain movements and the importance of flexible docking in virtual screening.
To illustrate the challenges of flexible docking, we re-trained the KarmaDock by integrating a ResNet module~\citep{he2016deep} (Supplemental Fig. S4) for predicting side chain torsion angles, resulting in a new model named KarmaDock-sc Align. DiffBindFR significantly surpasses KarmaDock-sc Align in terms of side chain recovery (Fig.~\ref{fig:3}.(B)). Compared to KarmaDock, the performance of KarmaDock-sc Align significantly declines (Fig.~\ref{fig:3}.(A)) due to the difficulty in balancing ligand coordinate recovery with side chain torsion recovery, highlighting the complexity inherent in flexible docking. As is widely recognized, factors like the number of heavy atoms and rotatable bonds in a ligand profoundly impact the success rate of conventional docking programs~\citep{pagadala2017software}. Hence, we examine the relationship between L-RMSD from the DiffBindFR model and various ligand characteristics, such as molecular weight, rotatable torsion bonds, and changes in solvent accessible surface area (DeltaSASA) upon binding. Contrary to traditional methods~\citep{friesner2004glide,trott2010autodock,ruiz2014rdock,Zhang2022ligpose}, as the ligand complexity and size increase, the performance of DiffBindFR-Smina does not deteriorate significantly (Fig.\ref{fig:3}.(C)). 

\subsection{Performance on the Posebusters test set}
Given that the similarity between samples to be predicted and those used in training can influence the performance of deep learning methods, Buttenschoen et al.~\citep{buttenschoen2023posebusters}, aiming for a more equitable comparison with traditional docking methods based on scoring functions, have curated a dataset called Posebusters test set from the PDB database. The Posebusters test set exclusively comprises 428 complexes on which the deep learning methods have not been trained. To evaluate the physical plausibility of poses generated by DiffBindFR, we compared its performance to other baseline methods using the Posebusters test set and the Posebusters suite~\citep{buttenschoen2023posebusters}, a tool designed to assess the validity of ligand-protein complexes based on criteria including bond length, planarity of aromatic rings in ligands, and clashes between ligands and proteins. Success in docking is redefined as a pose having an L-RMSD less than 2 \AA\ and simultaneously passing the physical validity check by Posebusters, with this success rate termed as the PB-success rate. The Posebusters test set comprises 428 distinct complexes released since 2021, with no overlap with the PDBbind v2020 dataset. To demonstrate that the success rate of DiffBindFR is not solely due to local ligand energy relaxation and its superior in side chain packing, the performance of other methods is evaluated with stricter energy minimization for the ligand, given the experimental side chain conformation. For poses generated by DiffBindFR, ligand energy minimization is conducted using the side chains as predicted by the model. The energy minimization is performed using the AMBER ff14sb force field~\citep{maier2015ff14sb} for prorein and the Sage force field~\citep{boothroyd2023development} for ligand in OpenMM~\citep{eastman2017openmm}, as used in the Posebusters paper~\citep{buttenschoen2023posebusters}. Fig.\ref{fig:4}.(A) shows that traditional rigid receptor docking methods like Glide perform best on re-docking when provided with the correct Holo pocket environment, followed by Vina and Gold, with most of their generated docking poses being physically valid. However, their performance significantly deteriorates when docking with Rosetta-repacked proteins, highlighting their heavy dependence on side chain conformations. Traditional flexible docking methods rDock and VinaFlex are also involved in comparison. VinaFlex, heavily reliant on predefined flexible side chains, performs the worst in our scenario where information about flexible side chains is assumed unavailable. rDock, capable of optimizing functional groups prone to forming hydrogen bonds in side chains, achieves higher success rate in repacked proteins compared to Vina and Glide, but lower success rate in proteins with ground-truth side chains. For these traditional methods, their PB-success rate is only slightly lower than their overall success rate, indicating that most generated poses is validated by Posebusters suite due to the physical components in their scoring functions. Therefore, the post ligand optimization using force field does not cause obvious impact to their PB-success rate.

\begin{table*}[h]
\scriptsize
\begin{center}   
\caption{Success rate of various methods on CD cross-dock test set.$^\alpha$}
\label{table:1}
\resizebox{\textwidth}{!}{
\begin{tabular}{cccccccc}  
\midrule
\multicolumn{2}{c}{\multirow{2}{*}{\scalebox{1.2}{Method}}} & \multicolumn{6}{c}{\scalebox{1.2}{PB-success rate}}\\
\cmidrule(lr){3-8}
& & Ensemble-CDK2 & Ensemble-EGFR &  Ensemble-FXA & ApoRef & CASF2016 & GPCR-AF2\\
\midrule
\multirow{5}{2.1cm}{Traditional rigid \\ receptor docking \\ methods} & Vina & 0.079 & 0.060 & 0.344 & 0.082 & 0.294 & 0.136\\
& LinF9 & 0.061 & 0.119 & 0.365 & 0.089 & 0.272 & 0.076\\
& Smina & 0.079 & 0.090 & 0.362 & 0.077 & 0.304 & 0.152\\
& Gnina & 0.099 & 0.090 & 0.394 & 0.082 & 0.320 & 0.106\\
& Glide & 0.154 & 0.090 & 0.271 & 0.091 & 0.219 & 0.182\\
\midrule
\multirow{2}{2.1cm}{Traditional flexible \\ docking methods} & VinaFlex & 0.013 & 0.000 & 0.005 & 0.015 & 0.023 & 0.045\\
& rDock & 0.257 & 0.134 & 0.440 & 0.157 & 0.296 & 0.212\\
\midrule
\multirow{5}{2.1cm}{Deep learning-based \\ docking methods} & TankBind-pocket & 0.100 & 0.015 & 0.110 & 0.040 & 0.123 & 0.015\\
& EDM-Dock & 0.051 & 0.015 & 0.009 & 0.011 & 0.064 & 0.00\\
& KarmaDock Align & 0.135 & 0.045 & 0.009 & 0.047 & 0.136 & 0.015\\
& DiffBindFR-Smina & 0.564 & 0.403 & 0.789 & 0.434 & 0.566 & \pmb{0.227}\\
& DiffBindFR-MDN & \pmb{0.674} & \pmb{0.478} & \pmb{0.794} & \pmb{0.476} & \pmb{0.636} & 0.212\\
\midrule
\end{tabular}}
\end{center}
\footnotesize{$^\alpha$Best performance in bold for the highest PB-success rate.}\\
\end{table*}

Among blind docking methods, DiffDock shows better performance (success rate of 38$\%$) than TankBind (16$\%$) and EquiBind (2$\%$), but most of their generated poses are invalid due to ignoring protein side chains (Fig.\ref{fig:4}.(A)). Ligand energy minimization significantly improves the PB-success rate of DiffDock (35$\%$). TankBind and EquiBind also see improvements in PB-success rate with energy minimization, but still lag behind DiffDock (Fig.\ref{fig:4}.(B)). 
Although blind docking is a tough task for its broad searching space in the whole protein, flexible pocket docking method like DiffBindFR, denoising a chaotic side chain conformation into a well-packed conformation having valid interaction with the ligand, has much more objectives for prediction. DiffBindFR, utilizing Smina scoring function or MDN network as the confidence model to select the top-1 pose from 40 generated ones, outperforms all other deep learning-based blind docking methods and pocket docking methods. DiffBindFR-Smina and DiffBindFR-MDN demonstrate both high success rate (50.2$\%$ for DiffBindFR-Smina and 48.1$\%$ for DiffBindFR-MDN) and PB-success rate (49.1$\%$ for DiffBindFR-Smina and 44.4$\%$ for DiffBindFR-MDN), with lower penalties by Posebusters compared to other deep learning-based methods, showcasing the capability of DiffBindFR in generating accurate and physically plausible complex poses. The performance of DiffBindFR is comparable to traditional rigid receptor docking methods using known ground-truth side chain conformations for redocking. As is depicted from Supplemental Fig. S7, DiffBindFR shows its effectiveness in binding site identification and pocket side chain recovery on Posebusters test set, as well. Force field optimization has minimal impact on DiffBindFR generated structures, which also demonstrates the high physical plausibility of DiffBindFR generated poses. Among other deep learning-based pocket docking methods, KarmaDock Align achieves the highest success rate (30.4$\%$) but a very low PB-success rate (6.1$\%$). Force field optimization of ligands rescues most poses with L-RMSD $\textless$ 2 \AA\ into good physical validity. KarmaDock Align, EDM-Dock, TankBind-pocket, Uni-Mol, and EDM-Dock, which focus on fitting the RMSD of the ligand during training, tend to ignore the intra energy of the generated poses and protein side chains, as is shown from supplemental Fig. S9. Indeed, force field optimization is not allowed in realistic docking to meet the demands of high-throughput screening.

Four specific cases from the Posebusters test set (supplementary Fig. S10), never trained or seen by DiffBindFR, are presented to highlight its superiority over other methods focusing solely on ligand coordinates while neglecting ligand conformation validity. In complexes with PDB ID 6TW5, 7PRM, 7T1D, and 7CD9, DiffBindFR successfully docks ligands into precise positions with valid conformations and recovers pocket side chains into good interaction with ligands. In contrast, KarmaDock Align, EDM-Dock, and TankBind-pocket fail to predict correct binding ligand poses, and their generated poses cannot pass the physical plausibility check of the Posebusters suite. As is shown in Supplementary Table. S4, ligand poses generated by EDM-Dock and TankBind-pocket exhibit both internal invalidity (including internal steric clash, bump aromatic ring, etc) and steric clash with proteins, while KarmaDock Align, due to using RDKit for ligand pose alignment, frequently fails in reducing ligand-protein clash.


\begin{figure*}[!h]
\centering
\includegraphics[scale=0.22]{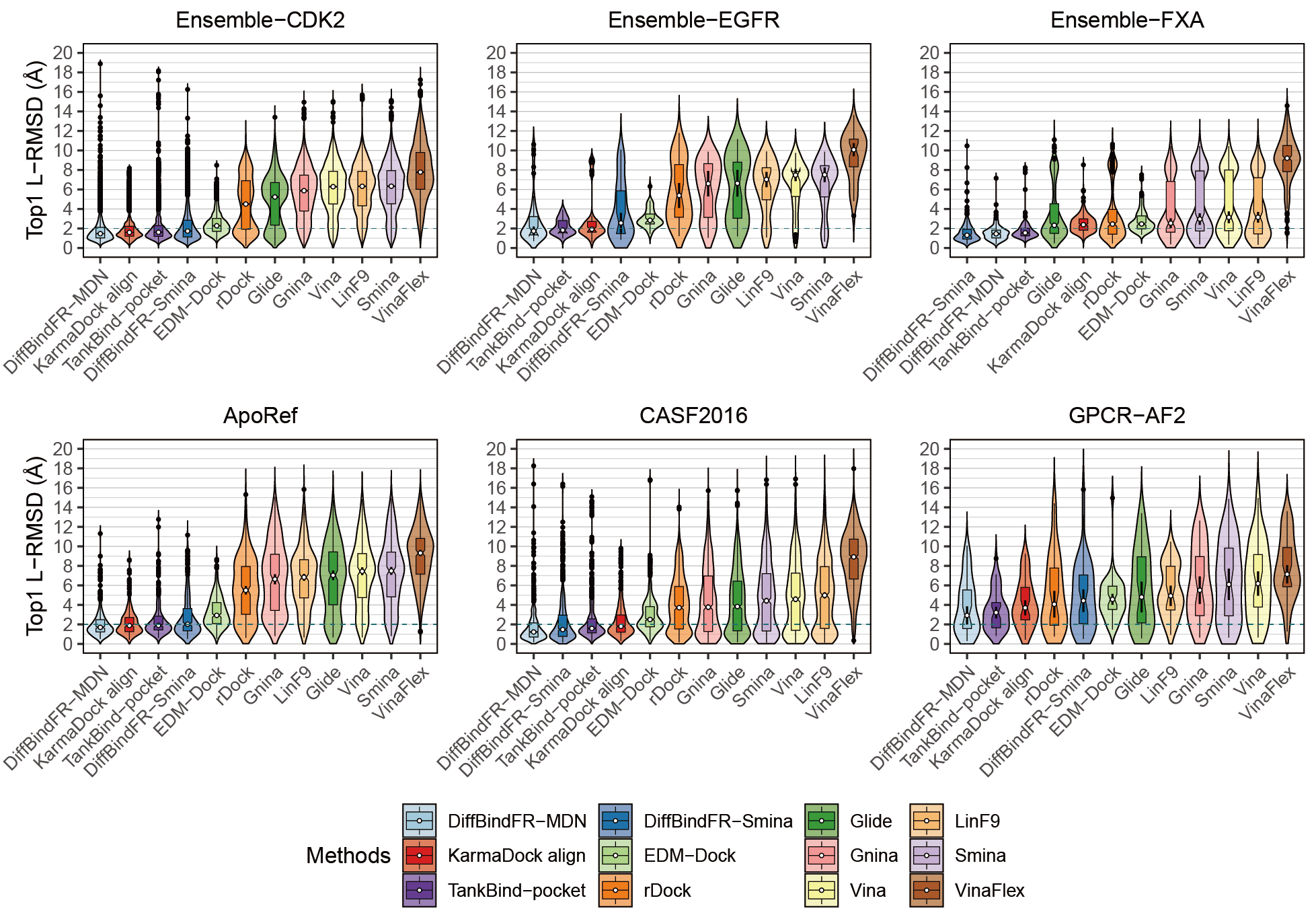}
\caption{The L-RMSD distribution of various methods on CD cross-dock test set.}

\label{fig:5}
\end{figure*}

\subsection{Performance on the CD cross-dock test set}
To showcase the exceptional capabilities of DiffBindFR in flexible docking, we evaluate its performance in the more challenging task of cross-docking. We use a self-curated benchmark called the CD test set, which includes various cross-docking scenarios such as Apo-Holo and cross-docking between different Holo states with various protein families. CD test set contains six subsets, ApoRef~\citep{zhang2022holo}, CASF2016~\citep{liu2017forging} with target proteins in the Apo state, GPCR-AF2 set with Apo-like proteins predicted by AlphaFold2~\citep{jumper2021highly}, and Ensemble sets featuring prominent docking targets including CDK2, EGFR and FXA. C$\mathrm{\alpha}$ RMSD of binding site backbone (within 5 \AA\ cutoff away from crystal ligand) conformational changes in these subsets predominantly range between 0-2 \AA, as shown in Supplemental Fig. S1.

\begin{figure*}[!h]
\centering
\includegraphics[scale=0.16]{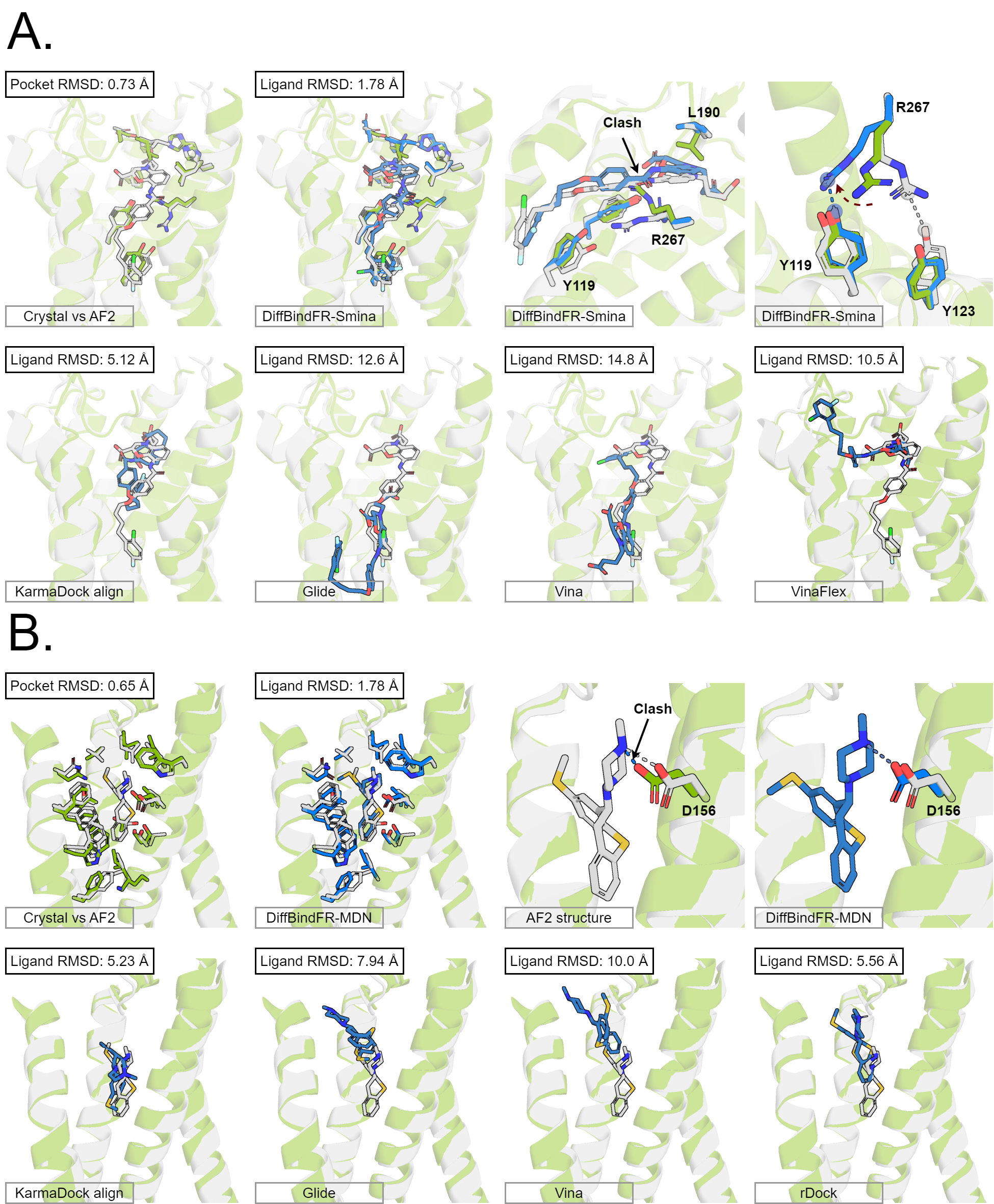}
\caption{The binding poses of two cases from the GPCR-AF2 subset in the CD test set. In all panels, Holo protein and ligand are shown in grey. AF2 modeled protein structure is shown in pale green. DiffBindFR sampled ligand and pocket side chains are shown in blue. Note that DiffBindFR calculations were done using the AF2 modeled backbone structures. For each frame, the docking method used is given at the bottom left. The pocket C$\mathrm{\alpha}$ RMSD between Holo and AF2 structure within 5 \AA\ cutoff away from crystal ligand, and ligand RMSD are reported on left top, except the two right top frames in each panel for the comparison between AF2 modeled and DiffBindFR sampled side chain conformations. (A). Human cysteinyl leukotriene receptor 2 bound to its antagonist ONO-2570366 (PDB ID: 6RZ6). The side chains of Y119, Y123, L190 and R267 are shown; (B). 5-Hydroxytryptamine receptor 2A bound to its inverse agonist methiothepin (PDB ID: 6WH4). The side chain of D155 is shown.}

\label{fig:6}
\end{figure*}

In these subsets, DiffBindFR-MDN and DiffBindFR-Smina achieve significantly higher PB-success rate (Table.\ref{table:1}) than all the traditional docking methods and deep learning-based docking methods. When considering L-RMSD alone (Fig.\ref{fig:5}), traditional rigid receptor docking methods such as Vina, Smina, LinF9~\citep{yang2021lin_f9}, and Glide underperform, were compared to deep learning-based methods that use main-chain coarse-grained representations of proteins. As indicated in Fig.\ref{fig:5} and Supplementary Table. S5, the L-RMSD median for methods like TankBind-pocket, EDM-Dock, and KarmaDock Align hovers around 2 \AA\ across the subsets, whereas for traditional rigid receptor docking methods, it even surpasses 5 \AA\ in subsets like CDK2, EGFR, and ApoRef.
However, when physical plausibility is taken into account, the PB-success rate for TankBind-pocket, EDM-Dock, and KarmaDock Align drops to levels similar to traditional rigid receptor docking methods (below 10$\%$). Notably, in the Ensemble-FXA subset, methods such as Vina, LinF9, Smina, Gnina, and Glide perform significantly better, achieving PB-success rates of 34.4$\%$, 36.5$\%$, 36.2$\%$, 39.4$\%$, and 27.1$\%$, respectively. Traditional flexible methods such as VinaFlex and rDock, developed for cross-dock scenarios, were also evaluated. VinaFlex shows poorer performance than rigid receptor docking methods in both L-RMSD distribution and PB-success rate, due to its reliance on predefined side chains and limitations on the number of flexible side chains. The flexible method rDock, capable of optimizing side chains conducive to hydrogen bonding, outperforms all traditional rigid receptor docking methods in L-RMSD distribution and has higher PB-success rate than TankBind-pocket, EDM-Dock, and KarmaDock Align. We observed that both traditional rigid receptor and flexible docking methods perform better in subsets like Ensemble-FXA and CASF2016, where pocket backbone conformational changes are minimal (mostly between 0-0.5 \AA, as is depicted from Supplementary Fig. S1). Our method DiffBindFR, leveraging a full-atom based neural network to learn additional side chain movements, marginally outperforms all other methods in accurately recalling ligand coordinates and ensuring the validity of complex poses. On the CD test set, the MDN network surpasses the Smina scoring function for pose ranking. DiffBindFR-MDN achieves state-of-the-art PB-success rate of 67.4$\%$, 47.8$\%$, 79.4$\%$, 47.6$\%$ and 63.6$\%$ in CDK2, EGFR, FXA, ApoRef and CASF2016, respectively. Futhermore, to investigate the efficacy of various methods in the context of cross-docking scenarios involving Apo-Holo pairs, we conduct a detailed computational assessment of these methods using a subset of 660 Apo-Holo pairs from the CASF2016 dataset. In this analysis, DiffBindFR-MDN emerges as the most effective technique, demonstrating superior performance in both terms of L-RMSD distribution and PB-success rate (56.1$\%$, as is shown from Supplementary Table. S6). Additionally, in the scenario as approximations of Apo-Holo cross-docking where the target proteins are predicted by AlphaFold2, specifically on the GPCR-AF2 subset, DiffBindFR-MDN exhibits slightly lower PB-success rate of 21.7$\%$ compared to DiffBindFR-Smina (22.7$\%$). This reduced performance is attributable to higher penalties incurred from ligand-protein clashes.

Here, we present the docking poses of various methods on two examples from the GPCR-AF2 subset (Fig.\ref{fig:6}). The first example involves a crystal structure with the PDB ID 6RZ6, where we investigate the target protein from 6RZ6 identified as the human Cysteinyl leukotriene receptor 2~\citep{gusach2019structural}. This receptor plays a role in regulating pro-inflammatory responses associated with allergic disorders. The ligand in this case is an antagonist, ONO-257036. We predict the AF2 structure (Apo-like) of the receptor protein and then dock the ligand molecule into this AF2 structure. Following binding sites alignment, the predicted AF2 structure exhibited a binding site (within 5 \AA\ cutoff away from crystal ligand) RMSD of 0.73 \AA\ when compared with the crystal structure. However, in the Apo-like AF2 predicted structure, residues R267 and L190 is found to block ONO-25703 binding to pocket, preventing traditional rigid receptor docking methods like Glide and Vina from locating the correct binding site. VinaFlex, despite its ability to leverage side chain flexibility, also fails to dock the ligand correctly. KarmaDock Align, similarly did not achieve correct docking, although it shows better L-RMSD than the methods mentioned above. In contrast, our flexible docking method, DiffBindFR, coupled with Smina and MDN confidence model can select the top-1 complex pose from the 40 poses. DiffBindFR adeptly repacks the side chain of the R267 residue, enabling the ligand to successfully dock into the correct binding position. The position of the R267 side chain predicted by DiffBindFR is somewhat different. In the crystal structure, the R267 side chain forms a hydrogen bond interaction with the Y123 side chain, whereas in the DiffBindFR-predicted structure, R267 forms a hydrogen bond interaction with the Y119 side chain, which can be attributed to the similarity in the spatial positions of the Y123 residue and the Y119 residue relative to R267.

The second example involves a crystal structure with the PDB ID 6WH4, where the target protein is the human 5-HT2A serotonin receptor, which is associated with the actions mediated by psychedelics~\citep{kim2020structure}, and the ligand is methiothepin. Following alignment of the binding sites, the AF2-predicted structure displays a pocket RMSD of 0.65 \AA\ when compared to the crystal structure. In the AF2 structure, the side chain of the D155 residue has VdW (Van der Waals) clash (pair distance is 1.5 \AA\ ) with the ligand, leading to the failure of docking attempts by Glide and Vina. Although KarmaDock Align and the traditional flexible docking method rDock predicts the ligand position with a smaller L-RMSD, they still do not achieve successful docking. In contrast, the top-1 complex pose selected by the MDN confidence model in DiffBindFR not only accurately reproduces the ligand position but also successfully repacks the side chain of D155 residue, enabling it to form electrostatic interactions with the ligand. This example further demonstrates the capability of DiffBindFR to effectively manage protein-ligand interactions, particularly in challenging docking scenarios.

We also present four cases from ApoRef subset where DiffBindFR successfully docks ligands into the Apo pockets of crystal structures (Supplementary Fig. S11), with these complexes not having appeared in the training set. The PDB IDs for these four cross-dock examples are as follows: Holo: 1ZGY, Apo: 1PRG; Holo: 2XIR, Apo: 1VR2; Holo: 3UVR, Apo: 1WFC; and Holo: 3RM6, Apo: 4EK3. In each of these instances, top-1 complex poses generated by DiffBindFR accurately recover the ligand poses, while side chains in the Apo state pockets that would otherwise clash with the ligand are also optimized.
These results highlight the potential of DiffBindFR in aiding researchers to study detailed interactions in real scenarios when no complex structures are available and provide insights for further lead optimization.

\section{Conclusions}
In this research, we have developed a full-atom diffusion model, DiffBindFR, for flexible pocket docking. DiffBindFR is capable of explicitly simulating the interactions of full atoms between the pocket side chains and the ligand molecules, which is extremely hard for previous docking methodologies. Our method not only surpasses traditional approaches in terms of the docking success rate but also achieves state-of-the-art (SOTA) levels in generating plausible docking conformations when compared to recent deep learning methods, as evidenced by evaluations conducted on the PDBbind and Posebusters test sets. Furthermore, starting from a random side chain conformation, DiffBindFR can accurately dock molecules while concurrently recovering the side chain conformations.

On the cross-docking benchmark, CD test set, DiffBindFR has also demonstrated superior performance. Notably, previous methods that employ deep learning to characterize protein pockets through coarse-grained main-chain representations also show promise results, but they lead to a lack of detailed information regarding the interactions between side chain atoms and ligands. DiffBindFR that simulates side chains addresses this gap in deep learning methodologies. Previous research~\citep{gaudreault2012side,clark2019inherent,wankowicz2022ligand} has indicated that the majority of proteins undergo minimal backbone alterations upon ligand binding, with the primary conformational changes caused by side chains. Therefore, DiffBindFR remains adept at predicting accurate docking poses in most scenarios with slight protein backbone movements. However, in cases where the pocket backbone exhibits significant flexibility (such as cryptic sites~\citep{meller2023predicting}), the docking results may not be as satisfactory. Recent benchmarking studies~\citep{diaz2023deep,zhang2023benchmarking,kersten2023hic,scardino2023good,holcomb2023evaluation,karelina2023accurately} have revealed that structures modeled by AF2 tend to exhibit a pocket backbone conformation more akin to the Holo state, presumably owing to the conservative functional sites within the multiple sequence alignment (MSA). However, inaccuracies in side chain placement often lead to suboptimal virtual screening performance when compared with the Holo pocket. In this context, DiffBindFR emerges as a promising tool for refining the side chains in AF2 modeled structures, potentially enhancing enrichment in virtual screening campaigns in the absence of available Holo structures. Further investigation of this application will be in future research endeavors.

The physical validity of DiffBindFR generated complex poses, coupled with the simulated detailed three-dimensional interactions, provides users with correct interactions to facilitate further optimization. In addition, the conformation alterations predicted by DiffBindFR will significantly augment comprehension of the molecular mechanisms underlying specific actions, and better elucidate the relationship between structure and function. 

\section{Methods}
\subsection{Data representation}
The complete set of heavy atoms from the ligand molecule and protein pocket is structured into a heterogeneous graph $G=(\mathcal{V},\mathcal{E})$, where each atom corresponds to a node. For the node representation $\mathcal{V}_p$ of pocket residue atoms, we employ one-hot encoding encompassing atom type, residue type, and whether the atom is part of the backbone. The ligand node features $\mathcal{V}_l$ include atom type, hybridization type, atomic connectivity, explicit valence, implicit valence, number of rings it belonging to, aromaticity, formal charge, partial charge, chirality type, the number of radical electrons, the number of hydrogens, and whether it is in an N-membered ring (with nitrogen ranging from 3 to 8). Furthermore, pharmacophore features such as hydrogen bond acceptor/donor, aromaticity, hydrophobicity, and positive/negative charge are integrated. The edges $\mathrm{e_{ij}}$, based on the covalent bonds of ligand, incorporate features like bond type, stereochemistry, conjugation, and whether the bond is part of a ring system. Diffusion times $\mathrm{t}$ are encoded using a sinusoidal format and are concatenated to the scalar representations of nodes and edges. For ligand atoms, internal edges $\mathcal{E}_{ll}$ connected by covalent bonds are pre-constructed. For pocket atoms, in addition to covalent bonds, edges $\mathcal{E}_{pp}$ are linked between pocket atoms and their own $C_\alpha$ and $C_\beta$ atoms. During the forward inference of the model, edges are dynamically constructed based on the three-dimensional coordinates of all atoms. Within the ligand molecule, the graph construction uses a cutoff radius of 5 \AA, and a similar cutoff is applied for the full atom graph of the pocket and directed edges from pocket to ligand atoms $\mathcal{E}_{pl}$. These edges, serving as non-covalent interactions, solely encode distances. Given the model's need to predict ligand translational updates, it's essential for the ligand to be aware of the entire pocket atom's position. Therefore, directed edges $\mathcal{E}_{lp}$ from ligand to pocket atoms are dynamically constructed based on the diffusion process, with the translational noise determining the cutoff radius as $\mathrm{0.2\sigma_{T}+5}$ \AA. This ensures that even in high noise scenarios, where the ligand is distant from the pocket, there remains an informational interaction between the ligand and the pocket, thereby pulling the ligand closer. All edges from the heterogeneous graph are $\{\mathcal{E}_{ll},\mathcal{E}_{pp},\mathcal{E}_{pl},\mathcal{E}_{lp}\}$, and their distance features utilize Gaussian radial basis for encoding.

\subsection{DiffBindFR score network}
The architecture of DiffBindFR is meticulously crafted upon the foundation provided by the e3nn library~\citep{geiger2022e3nn}. It primarily comprises the following pivotal components: a module for input embedding, modules for intra-molecular interaction encoding, and modules dedicated to inter-molecular interaction encoding. The network ingests a geometric heterogeneous graph, encompassing invariant scalar representations of both ligand heavy atom nodes $\mathcal{V}_l$ and pocket residue atom nodes $\mathcal{V}_p$. We harness the irreducible representations (Irreps) to encode features by spherical harmonics. As the depth of feature encoding advances, the scalar inputs evolve towards higher-order physical quantity representations. Every interaction module is constructed using the Tensor Product Layer (TPL), establishing SE(3) equivariant message-passing functions. The tensor products are realized by encoding edge vectors with spherical harmonic functions and then doing spherical tensor product of irreps with path weights. The weights of these tensor products are derived from a transformation of node representations constituting the edge and the edge representation itself through a layer of Multi-Layer Perceptrons (MLP); these weights also constitute the primary learnable parameters at each layer. For any given submodule, the general formula for message passing to node $a$ is:


\begin{equation}
\mathrm{h_a \xleftarrow{} h_a \bigoplus_{z \in \{l,p\}} LN^{(z_a,z)}(\frac{1}{|{\mathbf{N}_a}^{(z)}|} \sum_{b \in {\mathbf{N}_a}^{(z)}}Y(\hat{r}_{ab}) \bigotimes \psi_{ab}\ h_b)}
\end{equation}

$\mathrm{h_a} = (\mathrm{h_a^0, \overset{\rightarrow}{h_a}})$ represents Irreps of node $\mathrm{a}$, which is the concatenation of scalar representation $\mathrm{h_a^0}$ and vector representation $\mathrm{\overset{\rightarrow}{h_a}}$. $\mathrm{z_a}$ is the node type of node $\mathrm{a}$, and $\mathrm{z}$ can be any node type from the pocket node or the ligand node. $\mathbf{N}_a^{(z)}$ denotes the neighbour nodes of node $\mathrm{a}$. $\mathrm{Y}$ are the spherical harmonics up to $l=2$. $\mathrm{LN}$ is the equivariant layer normalization. $\mathrm{\bigoplus}$ refers to normal vector addition, and $\mathrm{\bigotimes \psi_{ab}}$ refers to the spherical tensor product of Irreps with path weights, with $\mathrm{\psi_{ab} = MLP^{(z_a,z)}(e_{ab},h_a^0,h_b^0)}$ following the graph message passing paradigm.

For predicting the scores of ligand translation and rotation, we construct a node $\mathrm{o}$ for each ligand center and gather the message from other ligand nodes to the center. We output the final single odd and single even vectors through layer normalization for translational and rotational score prediction as follows:
\begin{equation}
\begin{split}
\mathrm{[h_l^{(1o)}, h_l^{(1e)}] \xleftarrow{} \frac{1}{|\mathcal{V}_l|}\sum_{a \in \mathcal{V}_l} Y(\hat{r}_{oa})\bigotimes \psi_{oa}h_a},
\\ \mathrm{with\ \psi_{oa} = MLP(\mu_{oa}, h_a^0)}
\end{split}
\end{equation}
$\mathrm{\mu_{oa}}$ denotes the Gaussian radial embeddings for distance $\mathrm{r_{oa}}$. $\mathrm{h_l^{(1o)}}$ is the predicted score for translation. $\mathrm{h_l^{(1e)}}$ is the predicted score for rotation axis $\mathrm{\hat{\omega}}$.

For both the rotatable bonds of ligands and the dihedral angles of protein residue side chains, updates for each angle are anticipated based on a consistent paradigm. We define the central axis of the rotatable bond or dihedral angle as $\mathrm{B=(i,j)}$, represented by a bond formed by atoms $\mathrm{i}$ and $\mathrm{j}$. Further, the center of the rotatable bond is denoted as $\mathrm{c}$. A radius graph of ligand nodes with a 4 \AA\ cutoff is constructed to predict the torsion score of the ligand rotatable bonds.
\begin{equation}
\begin{split}
\mathrm{h_c \xleftarrow{} \frac{1}{\mathbf{N}_c} \sum_{a \in \mathbf{N}_c} Y^2(\hat{r}_{ab}) \bigotimes Y(\hat{r}_{ca}) \bigotimes  \psi_{ca} h_a},\\
\mathrm{with\ \psi_{ca}=MLP(\mu_{ca}, h_a, h_i+h_j)}
\end{split}
\end{equation}
To satisfy the parity of dihedral angles, spherical harmonics $\mathrm{Y^2}$ up to $l=2$ is utilized. We will employ the scalar features derived from the final obtained $\mathrm{h_c}$ to predict the torsion angles. An analogous procedure is adopted for the torsion of pocket side chains.

\subsection{Model training}
DiffBindFR neural network was trained using the AdamW optimizer~\citep{loshchilov2017decoupled} with a learning rate of 0.0005 and a batch size of 64 for 1000 epochs on eight 80 GB NVIDIA A800 TENSOR CORE GPUs. MDN confidence model was trained using the Adam optimizer~\citep{kingma2014adam} with a batch size of 256 for 1000 epochs on four 32 GB NVIDIA Tesla V100-SXM2 GPUs. 

\subsection{Baseline methods}
\subsubsection{Vina}
Autodock Vina~\citep{eberhardt2021autodock} is a widely-used traditional docking method. We defined the box using the center of the ligand present in the crystal structure, setting the box dimensions to 24×24×24 \AA$^3$. The 'exhaustiveness' parameter in Vina was set to 32, producing up to 10 poses for each docking run. Docking was repeated running 40 times with different random seeds to get the top-ranked pose.

\subsubsection{Smina}
Smina~\citep{koes2013lessons} improves Autodock Vina with a new scoring function and is more easy-to-use. The box construction and the sampling strategy were the same from the aforemetioned baseline method AutoDock Vina.

\subsubsection{LinF9}
LinF9~\citep{yang2021lin_f9} improves Autodock Vina with a new scoring function and is more user-friendly. The box construction and the sampling strategy were the same from the aforemetioned baseline method AutoDock Vina.

\subsubsection{Gold}
Gold~\citep{jones1997development} is another widely-used traditional docking method. The binding sites were defined as pocket residues within radius 12.5 \AA\ around the crystal ligand. The settings used were rescore function ‘plp’, autoscale 10, and early termination off. The docking performance was taken from Buttenschoen et al. reported~\citep{buttenschoen2023posebusters}. 

\subsubsection{VinaFlex}
AutoDock Vina also supports flexible docking with movable side chains~\citep{ravindranath2015autodockfr}. However, it requires the explicit designation of the side chains allowed to move and can support up to 14 flexible residues. Before each docking attempt, we randomly selected up to 14 residues within the defined 24×24×24 \AA$^3$ box to act as the flexible residues. The 'exhaustiveness' parameter was set to 16. Each docking run generated up to 10 poses, and this docking process was repeated running 40 times using different random seeds to get the top-ranked pose.

\subsubsection{rDock}
rDock~\citep{ruiz2014rdock} is another traditional docking method. The box construction and the sampling strategy were the same from the aforemetioned baseline method AutoDock Vina. Otherwise, functional groups, specifically -OH and -NH3+, located within 3 \AA\ of the ligand on the pocket residues were allowed to move. Docking was repeated running 40
times with different random seeds to get the top-ranked pose.

\subsubsection{Glide}
Glide~\citep{friesner2004glide} is a powerful commercial docking method. The rigid receptor docking was executed using the Glide-SP docking method in the Schrodinger software suite. The system was protonated at pH=7.0 and energy minimization was performed on hydrogen atoms using the OPLS 2005 force field. For the generation of grid files, the parameter 'INNERBOX' was set to 15 and 'UTERBOX' was set to 30, with all other parameters as default. Each docking run produced a maximum of 10 poses, and the docking was repeated running 40 times to get the top-ranked pose.

\subsubsection{TankBind}
TankBind~\citep{lu2022tankbind} is a recently developed deep learning-based method. Instead of using the P2Rank prediction for pocket localization, the model utilizes the center of the ligand from the crystal structure, with all other parameters set to their default values. Since this method reconstructs ligand coordinates from the predicted distance matrix of complex, it can only generate a single pose for the ligand.

\subsubsection{EDM-Dock}
EDM-Dock~\citep{masters2023deep} is a deep learning-based method sharing similar algorithm with TankBind. The box was defined as a 22.5×22.5×22.5 \AA$^3$ cube. Extra energy minimization was performed for the single ligand pose predicted by EDM-Dock.

\subsubsection{KarmaDock}
KarmaDock~\citep{zhang2023efficient} is a recently developed deep learning-based regression model which predicts ligand coordinates directly in the euclidean. Following the protocol from the KarmaDock article~\citep{zhang2023efficient}, we reproduced its reported results on the CASF2016 test set (Supplemental Table. S3), showing that we successfully re-trained the original KarmaDock. For fair comparison with our model, we further re-trained KarmaDock using the PDBbind time-split training set without any artificial intervention. KarmaDock docking was run with its default parameters.

Additionally, we augmented the KarmaDock model with a ResNet module to predict the side chain torsion angles of the binding pocket, resulting in a refined model named KarmaDock-sc. 

\subsubsection{DiffDock}
DiffDock~\citep{corso2022diffdock} is a blind-docking method based on diffusion generative model. Although it's not fair to compare DiffDock with pocket-docking methods, we still evaluate its performance to reflect the defect of ignoring physical plausibility of these deep learning-based methods. Each generation of ligand poses was repeated running 40 times, and the generated poses were ranked by DiffDock confidence model. Again, the docking performance was taken from Buttenschoen et al. reported~\citep{buttenschoen2023posebusters}. 

\section{Data availability}
The protein-ligand complexes of PDBbind v2020 dataset were downloaded from \href{https://zenodo.org/records/6408497}{https://zenodo.org/records/6408497}. The protein-ligand complexes of Posebusters test set were downloaded from \href{https://zenodo.org/records/8278563}{https://zenodo.org/records/8278563}. The protein-ligand complexes of CD cross-dock test set will be publicly available as soon as possible.

\section{Code availability}
The source code of DiffBindFR will be publicly available at 
\href{https://github.com/HBioquant/DiffBindFR}{https://github.com/HBioquant/DiffBindFR} after our paper has been published.

\section{Author contributions}
J.Z and Z.G. designed the research, wrote source
code and performed the experiments. J.P. and L.L. designed and supervised the project. J.Z analyzed the experimental results. Z.G. and J.Z wrote the manuscript. J.P. and L.L. revised the manuscript. All authors read and approved the final manuscript.

\section{Conflicts of interest}
There are no conflicts to declare.

\section{Acknowledgements}
This work has been supported in part by the National Natural Science Foundation of China (22033001 and 32270689), the National Key R\&D Program of China (2023YFF1205103), the Chinese Academy of Medical Sciences (2021-I2M-5-014) and the Anhui's Plans for Major Provincial Science\&Technology Projects (202303a07020009). We thank the Computing Platform of the Center for Life Science (Peking University) for providing resources for the GPU-based model training. Part of the computation was performed on the computing platform of the Infinite Intelligence Pharma Ltd.



\balance


\bibliography{rsc} 
\bibliographystyle{rsc} 










\end{document}